\documentclass[letterpaper,twocolumn,english,pra, twocolumn,showpacs,superscriptaddress]{revtex4-1}

\usepackage[T1]{fontenc}
\usepackage{color}
\usepackage{babel}
\usepackage{pifont}
\usepackage{mathrsfs}
\usepackage{amsmath}
\usepackage{amssymb}
\usepackage{graphicx}
\usepackage[unicode=true,pdfusetitle,
 bookmarks=true,bookmarksnumbered=false,bookmarksopen=false,
 breaklinks=false,pdfborder={0 0 0},pdfborderstyle={},backref=false,colorlinks=true]
 {hyperref}
\hypersetup{
 urlcolor=blue, citecolor=blue}

\makeatletter

\pdfpageheight\paperheight
\pdfpagewidth\paperwidth

\AtBeginDocument{
  
}

\makeatother

\begin{document}

\title{Quantum statistics of a single atom Scovil\textendash Schulz-DuBois
heat engine}

\author{Sheng-Wen Li}

\address{Institute of Quantum Science and Engineering, Texas A\&M University,
College Station, TX 77843}

\affiliation{Baylor University, Waco, TX 76798}

\author{Moochan B. Kim}

\address{Institute of Quantum Science and Engineering, Texas A\&M University,
College Station, TX 77843}

\author{Girish S. Agarwal}

\address{Institute of Quantum Science and Engineering, Texas A\&M University,
College Station, TX 77843}

\author{Marlan O. Scully}

\address{Institute of Quantum Science and Engineering, Texas A\&M University,
College Station, TX 77843}

\affiliation{Baylor University, Waco, TX 76798}

\date{\today}
\begin{abstract}
We study the statistics of the lasing output from a single atom quantum
heat engine, which was originally proposed by Scovil and Schulz-DuBois
(SSDB). In this heat engine model, a single three-level atom is  coupled
with an optical cavity, and contacted with a hot and a cold heat bath
together. We derive a fully quantum laser equation for this heat engine
model, and obtain the photon number distribution for both below and
above the lasing threshold. With the increase of the hot bath temperature,
the population is inverted and lasing light comes out. However, we
notice that if the hot bath temperature keeps increasing, the atomic
decay rate is also enhanced, which weakens the lasing gain. As a result,
another critical point appears at a very high temperature of the hot
bath, after which the output light become thermal radiation again.
To avoid this double-threshold behavior, we introduce a four-level
heat engine model, where the atomic decay rate does not depend on
the hot bath temperature. In this case, the lasing threshold is much
easier to achieve, and the double-threshold behavior disappears.
\end{abstract}
\maketitle

\section{Introduction}

\begin{figure}
\includegraphics[width=0.5\columnwidth]{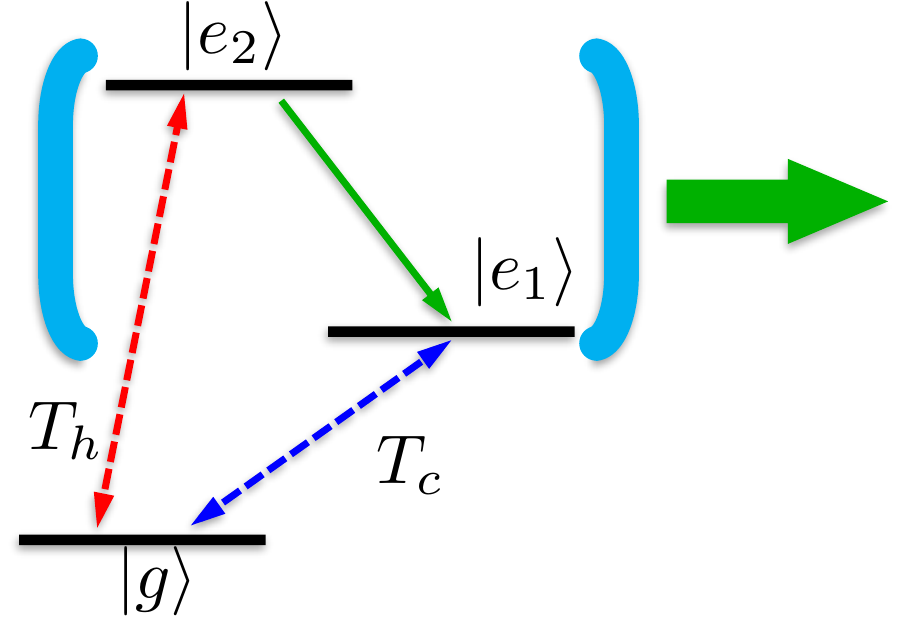}

\caption{(Color online) Demonstration for the SSDB heat engine. A three-level
atom is placed in an optical cavity to generate laser. We denote $\hbar\omega_{h}=E_{2}-E_{g}$,
$\hbar\omega_{c}=E_{1}-E_{g}$, and $\hbar\Omega_{l}=E_{2}-E_{1}$.}
\label{fig-3level}
\end{figure}

In 1959, Scovil and Schulz-DuBois introduced a quantum heat engine
model (SSDB heat engine) \cite{scovil_three-level_1959,geusic_quantum_1967},
where a single three-level atom is in contact with two heat baths
together (Fig.\,\ref{fig-3level}), and the population inversion
between the levels $|e_{1}\rangle$ and $|e_{2}\rangle$ can be created
by a large enough temperature difference giving rise to laser output.
During one working ``cycle'', one hot photon $\hbar\omega_{h}$
is absorbed, one cold photon $\hbar\omega_{c}$ is emitted, and one
laser photon $\hbar\Omega_{l}$ is produced. Thus, they obtain the
efficiency of the heat engine as $\eta_{\text{\textsc{ssdb}}}:=\Omega_{l}/\omega_{h}$.
To guarantee the laser output, a population inversion condition is
required $\exp(-\frac{\omega_{h}}{T_{h}})\ge\exp(-\frac{\omega_{c}}{T_{c}})$,
which is obtained from the considerations of counting the Boltzmann
factors. That simply leads to an upper bound for the SSDB efficiency
$\eta_{\text{\textsc{ssdb}}}\le1-T_{c}/T_{h}$, which is just the
Carnot limit. And it turns out that the SSDB heat engine is deeply
connected with many other quantum heat engine models, e.g., the quantum
absorption refrigerator \cite{geva_three-level_1994,linden_how_2010,chen_quantum_2012,kosloff_quantum_2017},
the electromagnetically-induced-transparency (EIT) based  heat engine
\cite{harris_electromagnetically_2016,zou_quantum_2017}, and it also
can be used to describe the photosynthesis process and solar cell
\cite{scully_quantum_2011,su_photoelectric_2016}.

This heat engine model gives a simple and clear demonstration for
the quantum thermodynamics. But we notice that some detailed properties
of this lasing heat engine, e.g., the threshold behaviour and the
statistics of the output light, is still not well studied. In Ref.\,\cite{scully_quantum_2011},
a rate equation description has been developed. In order to obtain
the photon statistics, we need to go beyond the rate equation description.
In this paper, we study this SSDB heat engine based on a more realistic
single-atom lasing setup \cite{meschede_one-atom_1985,agarwal_exact_1986,agarwal_steady_1990,walther_single_2000,teuber_nonclassical_2015},
where the three-level atom is placed in an optical cavity, and  coupled
with the quantized field mode, as well as in contact with two heat
baths with temperatures $T_{h,c}$ \cite{boukobza_thermodynamics_2006,boukobza_thermodynamic_2006,boukobza_three-level_2007,rahav_heat_2012,perl_thermodynamic_2017,yuge_decomposition_2017,ansari_entropy_2017}.
We derive the lasing equation in both semi-classical and fully quantum
approaches (Scully-Lamb approach \cite{scully_quantum_1966,scully_quantum_1967,scully_quantum_1997}),
and analytically obtain the photon number distribution in the steady
state for both above and below threshold cases. 

Intuitively, a higher temperature $T_{h}$ from the hot bath enhances
the population inversion between the two levels $|e_{1}\rangle$ and
$|e_{2}\rangle$, and thus should also enhance lasing. However, our
analytical result shows that a higher temperature $T_{h}$ also increases
the atomic decay rate. As the result, the lasing gain decreases when
$T_{h}$ is too high, and this system shows a ``double-threshold''
behavior: when the hot bath temperature $T_{h}$ is quite low ($T_{h}\simeq T_{c}$),
the excitation is too weak and the system is below the lasing threshold;
with the increasing of $T_{h}$, population inversion happens and
the lasing light comes out; but when $T_{h}$ keeps increasing, the
lasing gain starts to decrease and even goes below the threshold again,
thus another critical point appears, after which the output light
becomes thermal radiation again.

To avoid this double-threshold behavior, we study a four-level model
where a third ancilla bath is introduced \cite{yu_four-level_2010}.
In this model, neither of the two lasing levels is coupled with the
hot bath directly, and thus the atomic decay rate no longer depends
on the hot bath temperature. As the result, the lasing gain and cavity
photon number increases monotonically and only one critical point
exists. And it turns out the laser output  of this four-level heat
engine is also bounded by the Carnot efficiency.

We arrange the paper as follows: in Sec.\,II we introduce our model
setup and give a semi-classical analysis; in Sec.\,III we study the
full quantum theory, and derive the laser master equation. The master
equation has the same structure as the Scully-Lamb master equations,
however, with gain, loss and saturation parameters specific to the
three-level model of Scovil and Schulz-DuBois. In Sec.\,IV, we present
results for the photon statistics, we note the unusual feature that
for a given gain, the photon distribution could be different. The
quantum statistical features of the four-level model are presented
in Sec.\,V. We conclude with a summary in Sec.\,VI. Detailed derivations
are relegated to the Appendices.

\section{The SSDB heat engine }

The heat engine model is demonstrated in Fig.\,\ref{fig-3level}
\cite{boukobza_thermodynamic_2006,boukobza_thermodynamics_2006,boukobza_three-level_2007,yuge_decomposition_2017}.
A three-level system, $\hat{H}_{0}=E_{g}|g\rangle\langle g|+E_{1}|e_{1}\rangle\langle e_{1}|+E_{2}|e_{2}\rangle\langle e_{2}|$,
is placed in an optical cavity which is resonant with the atomic transition
$|e_{1}\rangle\leftrightarrow|e_{2}\rangle$. The transition path
$|e_{1/2}\rangle\leftrightarrow|g\rangle$ is coupled with a cold/hot
bath.

We denote the atomic transition operators as $\hat{\tau}_{h}^{-}:=|g\rangle\langle e_{2}|,\,\hat{\tau}_{c}^{-}:=|g\rangle\langle e_{1}|$,
$\hat{\sigma}^{-}:=|e_{1}\rangle\langle e_{2}|$, and $\hat{\tau}_{i}^{+}:=(\hat{\tau}_{i}^{-})^{\dagger}$,
$\hat{\sigma}^{+}:=(\hat{\sigma}^{-})^{\dagger}$. The atom and the
cavity interact resonantly through the Jaynes-Cummings coupling $\hat{V}=g(\hat{\sigma}^{+}\hat{a}+\hat{\sigma}^{-}\hat{a}^{\dagger})$,
and the dynamics of this cavity-QED system can be described by the
following master equation (interaction picture), 
\begin{equation}
\dot{\rho}=i[\rho,\hat{V}]+{\cal L}_{h}[\rho]+{\cal L}_{c}[\rho]+{\cal L}_{\mathsf{cav}}[\rho],\label{eq:ME}
\end{equation}
where
\begin{align}
{\cal L}_{i}[\rho]= & \gamma_{i}\overline{\mathsf{n}}_{i}\big(\hat{\tau}_{i}^{+}\rho\hat{\tau}_{i}^{-}-\frac{1}{2}\{\hat{\tau}_{i}^{-}\hat{\tau}_{i}^{+},\rho\}\big)\nonumber \\
+ & \gamma_{i}(\overline{\mathsf{n}}_{i}+1)\big(\hat{\tau}_{i}^{-}\rho\hat{\tau}_{i}^{+}-\frac{1}{2}\{\hat{\tau}_{i}^{+}\hat{\tau}_{i},\rho\}\big),\quad i=h,c\nonumber \\
{\cal L}_{\mathsf{cav}}[\rho]= & \kappa\big(\hat{a}\rho\hat{a}^{\dagger}-\frac{1}{2}\hat{a}^{\dagger}\hat{a}\rho-\frac{1}{2}\rho\hat{a}^{\dagger}\hat{a}\big).
\end{align}
 ${\cal L}_{h/c}[\rho]$ is the contribution from the hot/cold bath
coupled with the atom, and ${\cal L}_{\mathsf{cav}}[\rho]$ describes
the light leaking from the cavity to the outside vacuum field. Here
$\overline{\mathsf{n}}_{i}:=\overline{\mathrm{n}}_{\mathrm{\textsc{p}}}(\omega_{i},T_{i})$
for $i=h,c$ is the thermal photon number of the hot/cold bath calculated
from the Planck distribution $\overline{\mathrm{n}}_{\mathrm{\textsc{p}}}(\omega,T):=[\exp(\hbar\omega/k_{\text{\textsc{b}}}T)-1]^{-1}$. 

With this master equation, we obtain the equations of motion 
\begin{align}
\frac{d}{dt}\langle\hat{\textsc{n}}_{1}\rangle & =\gamma_{c}\big[\overline{\mathsf{n}}_{c}\langle\hat{\textsc{n}}_{g}\rangle-(\overline{\mathsf{n}}_{c}+1)\langle\hat{\textsc{n}}_{1}\rangle\big]-ig[\langle\hat{\sigma}^{-}\hat{a}^{\dagger}\rangle-\mathbf{h.c.}],\nonumber \\
\frac{d}{dt}\langle\hat{\textsc{n}}_{2}\rangle & =\gamma_{h}\big[\overline{\mathsf{n}}_{h}\langle\hat{\textsc{n}}_{g}\rangle-(\overline{\mathsf{n}}_{h}+1)\langle\hat{\textsc{n}}_{2}\rangle\big]+ig[\langle\hat{\sigma}^{-}\hat{a}^{\dagger}\rangle-\mathbf{h.c.}],\nonumber \\
\frac{d}{dt}\langle\hat{\sigma}^{-}\rangle & =ig\langle\hat{\sigma}^{z}\hat{a}\rangle-\frac{1}{2}\boldsymbol{\varGamma}\langle\hat{\sigma}^{-}\rangle,\nonumber \\
\frac{d}{dt}\langle\hat{a}\rangle & =-\frac{\kappa}{2}\langle\hat{a}\rangle-ig\langle\hat{\sigma}^{-}\rangle,\label{eq:semi-classical}
\end{align}
 where we denote $\hat{\textsc{n}}_{g}:=|g\rangle\langle g|$, $\hat{\textsc{n}}_{1,2}:=|e_{1,2}\rangle\langle e_{1,2}|$
$\hat{\sigma}_{z}:=\hat{\textsc{n}}_{2}-\hat{\textsc{n}}_{1}$ for
the atom operators, and 
\begin{equation}
\boldsymbol{\varGamma}:=\gamma_{h}(\overline{\mathsf{n}}_{h}+1)+\gamma_{c}(\overline{\mathsf{n}}_{c}+1)\label{eq:Gamma}
\end{equation}
 for the atomic coherence decay rate. 

We apply the semi-classical approximation that $\langle\hat{\sigma}^{-}\hat{a}^{\dagger}\rangle\simeq\langle\hat{\sigma}^{-}\rangle\langle\hat{a}^{\dagger}\rangle$,
$\langle\hat{\sigma}^{z}\hat{a}\rangle\simeq\langle\hat{\sigma}^{z}\rangle\langle\hat{a}\rangle=\langle\hat{\textsc{n}}_{2}-\hat{\textsc{n}}_{1}\rangle\langle\hat{a}\rangle$,
and assume the atom rapidly decays to its steady state right before
the cavity evolves significantly. Thus the quantum coherence term
is given by $\langle\hat{\sigma}^{-}\rangle=(2ig/\boldsymbol{\varGamma})\langle\hat{\textsc{n}}_{2}-\hat{\textsc{n}}_{1}\rangle\langle\hat{a}\rangle$
(denoting ${\cal E}:=\langle\hat{a}\rangle$), which is proportional
to the population inversion $\Delta\mathrm{N}$: 
\begin{align}
\Delta\mathrm{N}: & =\langle\hat{\textsc{n}}_{2}-\hat{\textsc{n}}_{1}\rangle=\cfrac{\overline{\mathsf{n}}_{h}-\overline{\mathsf{n}}_{c}}{\boldsymbol{\Phi}+\frac{4g^{2}|{\cal E}|^{2}}{\boldsymbol{\varGamma}}\boldsymbol{\Psi}}\label{eq:dN}\\
\boldsymbol{\Psi}: & =\frac{1}{\gamma_{h}\gamma_{c}}[\gamma_{h}(3\overline{\mathsf{n}}_{h}+1)+\gamma_{c}(3\overline{\mathsf{n}}_{c}+1)],\nonumber \\
\boldsymbol{\Phi}: & =3\overline{\mathsf{n}}_{h}\overline{\mathsf{n}}_{c}+2(\overline{\mathsf{n}}_{h}+\overline{\mathsf{n}}_{c})+1.\nonumber 
\end{align}
 Notice that when there is no cavity coupling ($g=0$), the atomic
populations return to the SSDB result 
\begin{equation}
\langle\hat{\textsc{n}}_{g}\rangle:\langle\hat{\textsc{n}}_{1}\rangle:\langle\hat{\textsc{n}}_{2}\rangle=1:\frac{\overline{\mathsf{n}}_{c}}{\overline{\mathsf{n}}_{c}+1}:\frac{\overline{\mathsf{n}}_{h}}{\overline{\mathsf{n}}_{h}+1},
\end{equation}
and the population inversion is
\begin{equation}
\Delta\mathrm{N}_{0}=(\overline{\mathsf{n}}_{h}-\overline{\mathsf{n}}_{c})/\boldsymbol{\Phi}.
\end{equation}
We see the constant $\boldsymbol{\Phi}$ is just the normalization
factor. 

\begin{figure}
\includegraphics[width=1\columnwidth]{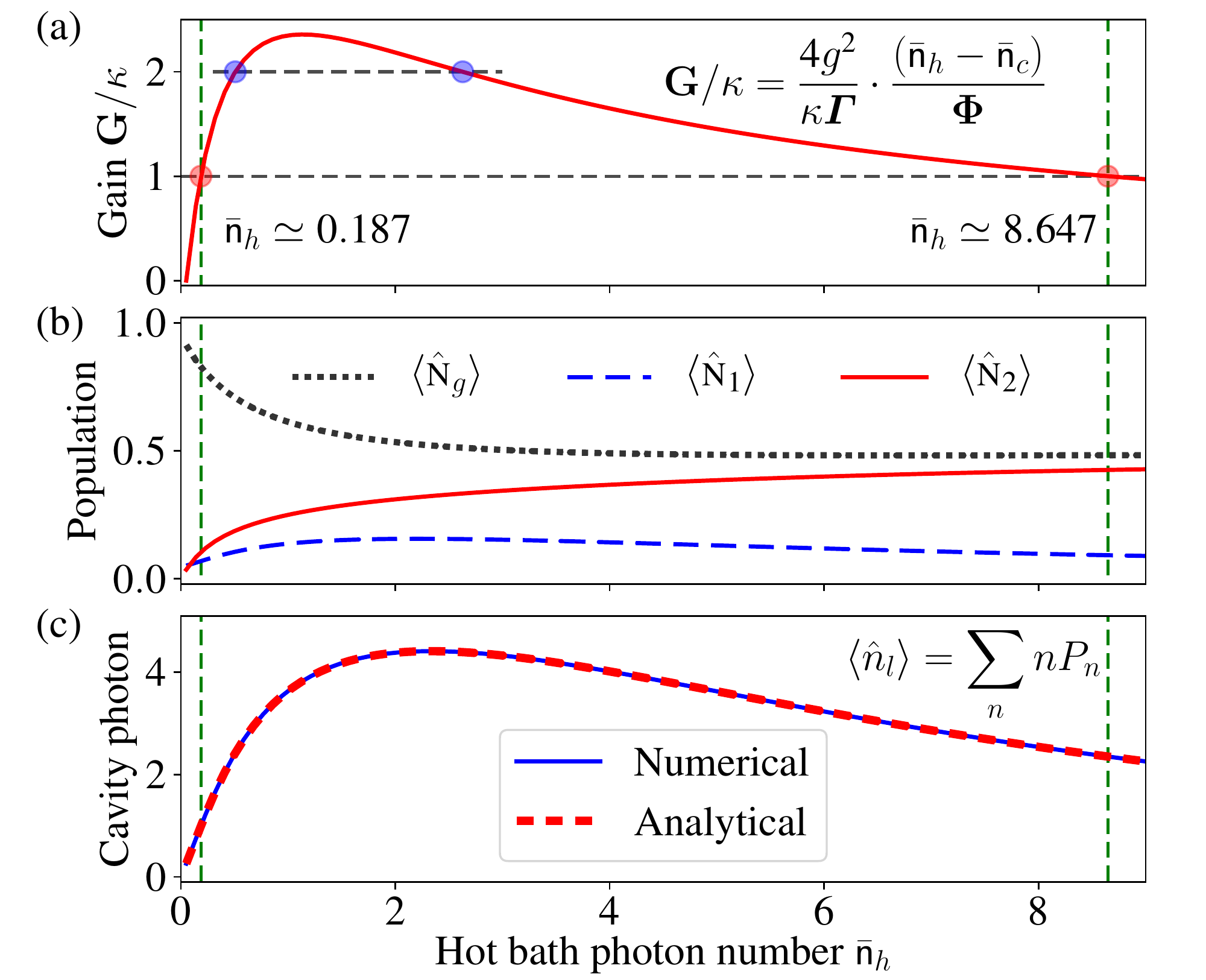}

\caption{(Color online) (a) The lasing gain $\mathbf{G}$. $\mathbf{G}/\kappa\ge1$
means above the lasing threshold. (b) The steady state populations
on $|g\rangle$, $|e_{1,2}\rangle$. (c) The average photon number
$\langle\hat{n}_{l}\rangle$ in the cavity obtained from the analytical
result Eqs.\,(\ref{eq:ratio}, \ref{eq:N_L}) (dashed red) and numerically
solving the master equation directly (solid blue). We set $\gamma_{h}=\gamma_{c}=32\kappa$,
$g=14\kappa$, and $\overline{\mathsf{n}}_{c}=0.05$ as the cold bath
photon number. The two critical points are $\overline{\mathsf{n}}_{h}\simeq0.187$
and $\overline{\mathsf{n}}_{h}\simeq8.647$. }

\label{fig-threshold}
\end{figure}

Now we obtain the lasing equation as
\begin{equation}
\dot{{\cal E}}=\big[\cfrac{2g^{2}(\overline{\mathsf{n}}_{h}-\overline{\mathsf{n}}_{c})}{\boldsymbol{\varGamma}\boldsymbol{\Phi}+4g^{2}|{\cal E}|^{2}\boldsymbol{\Psi}}-\frac{\kappa}{2}\big]{\cal E}=\frac{1}{2}\big[\frac{\mathbf{G}}{1+\mathbf{B}|{\cal E}|^{2}}-\kappa\big]{\cal E}.\label{eq:da/dt}
\end{equation}
 In the above bracket, $\mathbf{G}:=4g^{2}\Delta\mathrm{N}_{0}/\boldsymbol{\varGamma}$
is the lasing gain, and $\mathbf{G}/\kappa\ge1$ means above the lasing
threshold. And $\mathbf{B}:=4g^{2}\boldsymbol{\Psi}/\boldsymbol{\varGamma}\boldsymbol{\Phi}$
is the saturation parameter. It is worth noticing that, although the
population inversion $\Delta\mathrm{N}_{0}$ increases with $\overline{\mathsf{n}}_{h}$,
it also gets saturated and could never exceed 1, while the atomic
decay rate $\boldsymbol{\varGamma}$ keeps increasing linearly with
$\overline{\mathsf{n}}_{h}$.

As the result, with the increasing of $T_{h}$ starting from $T_{c}$,
the lasing gain first increases from zero, and gets above the threshold;
but then the lasing gain achieves a maximum point, after which it
starts to decrease, and even goes below the threshold again at a very
high temperature of $T_{h}$ {[}Fig.\,\ref{fig-threshold}(a, b){]}. 

Intuitively, a higher $T_{h}$ would enhance the population inversion
for lasing. But a higher $T_{h}$ also enhances the atomic decay rate
$\boldsymbol{\varGamma}$, and that suppresses the lasing gain {[}Eq.\,(\ref{eq:da/dt}){]}.
Therefore, at a very high temperature $T_{h}$, the lasing gain decreases
and even below the threshold again.

In Fig.\,\ref{fig-threshold}(b) we show a numerical result for the
atomic populations in the steady state changing with $T_{h}$. When
$T_{h}$ is very high, the populations on $|e_{1,2}\rangle$ have
been almost totally inverted, but the lasing gain $\mathbf{G}$ decreases
with $T_{h}$. As well, the cavity photon number $\langle\hat{n}_{l}\rangle$
shows the similar behavior {[}Fig.\,\ref{fig-threshold}(c){]}. Notice
that the photon number $\langle\hat{n}_{l}\rangle$ in the cavity
is not large, this is because we have only one atom in the cavity,
thus the photon emission is limited.

If the cavity coupling strength $g$ is strong, or atomic spontaneous
decay rates $\gamma_{h,c}$ are weak, the second critical point would
appear at a much higher temperature $T_{h}$, but such a behavior
of double critical points always exists. For realistic laser systems
with $N$ atoms in the cavity, the coupling strength could be effectively
enhanced by the atom number ($\sqrt{N}g$). Therefore, it is not easy
to observe such double-threshold behavior in common laser systems,
since the second threshold is usually too high and beyond the practical
regime of interests. However, for single atom heat engine laser, this
double-threshold behavior is much easier to happen.

In the fine cavity limit, $\kappa\rightarrow0$, this threshold condition
simply reduces as $\overline{\mathsf{n}}_{h}-\overline{\mathsf{n}}_{c}\ge0$,
and then it leads to the SSDB inequality $\eta_{\text{\textsc{ssdb}}}=\Omega_{l}/\omega_{h}\le1-T_{c}/T_{h}$,
which was derived based on the comparison of the Boltzmann factors
\cite{scovil_three-level_1959}. 

\section{Fully quantum approach}

The semi-classical approach is helpful to get a basic understanding
of the physics process in this heat engine. To get a more precise
and rigorous description, we adopt the Scully-Lamb approach to study
the fully quantum theory for the cavity mode $\varrho:=\mathrm{tr}_{\mathrm{atom}}\rho$
\cite{scully_quantum_1966,scully_quantum_1967,scully_quantum_1997,sargent_laser_1978}.
In this approach, the previous semi-classical separation of the correlation
functions are not needed. Denoting the matrix elements of $\varrho$
in Fock basis as $P_{mn}:=\langle m|\varrho|n\rangle$, we have
\begin{align}
\frac{d}{dt}P_{mn}= & ig\big(\sqrt{n}\rho_{12;m,n-1}-\sqrt{m}\rho_{21;m-1,n}\big)\nonumber \\
- & ig\big(\sqrt{m+1}\rho_{12;m+1,n}-\sqrt{n+1}\rho_{21;m,n+1}\big)\\
+ & \kappa\big[\sqrt{(m+1)(n+1)}P_{m+1,n+1}-\frac{1}{2}(m+n)P_{mn}\big].\nonumber 
\end{align}
Here $\rho_{\alpha\beta;mn}:=\langle\alpha,m|\rho|\beta,n\rangle$,
and $\alpha,\beta=1,2,g$ is the atom state indices. The first two
terms means the dynamics of the cavity mode and the atom are coupled
together, and we need to eliminate the atom degree of freedom.

\begin{figure}
\includegraphics[width=1\columnwidth]{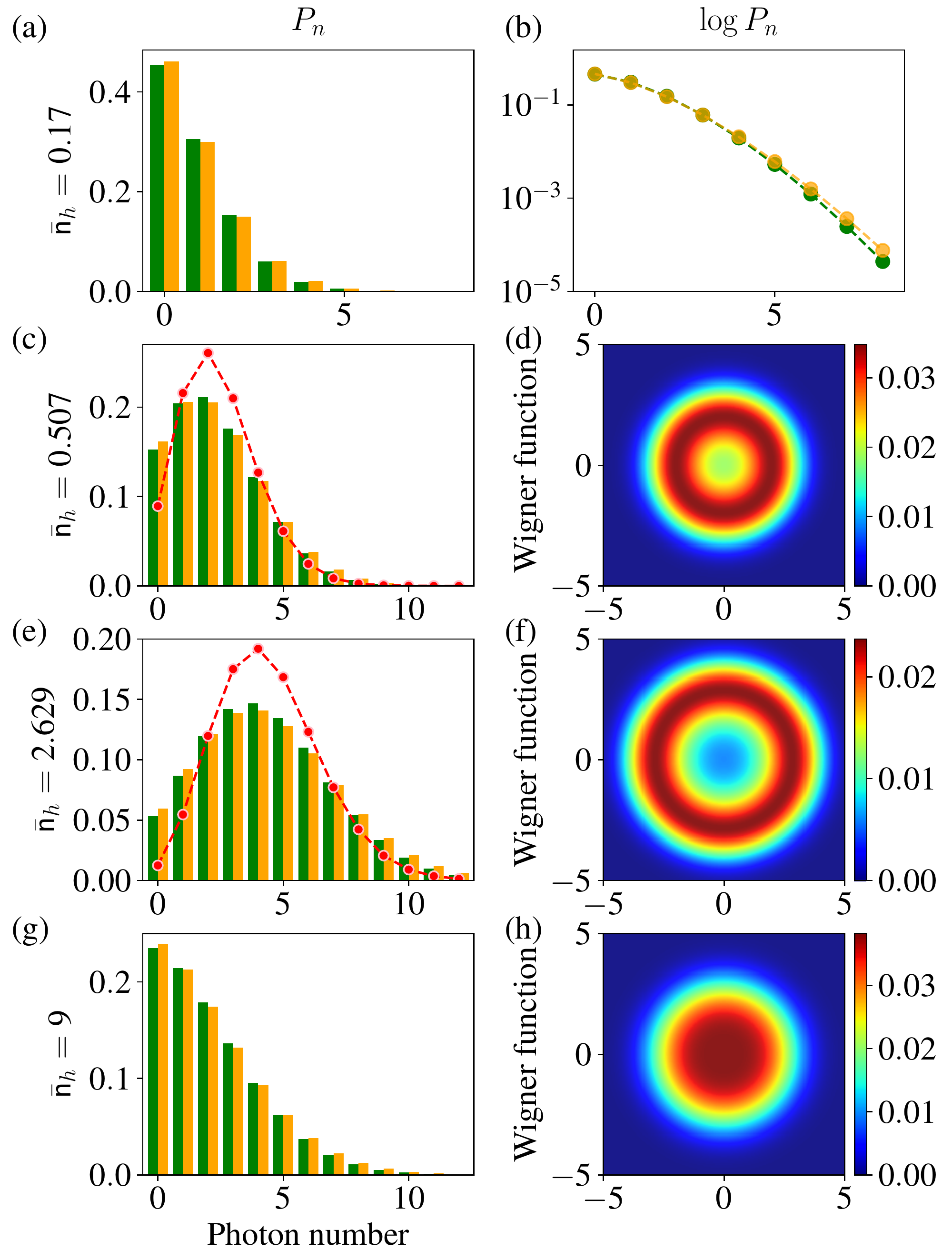}

\caption{(Color online) The photon number distributions and Wigner functions.
The parameters are the same with those in Fig.\,\ref{fig-threshold},
and (a, b) $\overline{\mathsf{n}}_{h}=0.17$ (below threshold) (c,
d) $\overline{\mathsf{n}}_{h}=0.507$ (above threshold) (e, f) $\overline{\mathsf{n}}_{h}=2.629$
(above threshold) (g, h) $\overline{\mathsf{n}}_{h}=9$ (below threshold).
Notice that $\overline{\mathsf{n}}_{h}=0.507$ (c, d) and $\overline{\mathsf{n}}_{h}=2.629$
(e, f) are the two blue points in Fig.\,\ref{fig-threshold}(a) which
have the same gain $\mathbf{G}/\kappa=2$. The two critical points
are $\overline{\mathsf{n}}_{h}\simeq0.187$ and $\overline{\mathsf{n}}_{h}\simeq8.647$.
The yellow columns (right) are given by the analytical result Eq.\,(\ref{eq:ratio}),
and the green columns (left) are the numerical result by solving the
master equation (\ref{eq:ME}) directly. The distributions below the
lasing threshold (a, b, g, h) are not exactly thermal distributions
($P_{n}\propto\exp[-n\Omega_{l}/T]$). The red dashed lines in (c,
e) are the corresponding Poisson distribution $P_{n}=e^{-\langle\hat{n}_{l}\rangle}\langle\hat{n}_{l}\rangle^{n}/n!$
where $\langle\hat{n}_{l}\rangle$ is the average photon number. }

\label{fig-wigner}
\end{figure}

For this purpose, we adopt the adiabatic elimination to take away
the dynamics of the atom \cite{scully_quantum_1966,scully_quantum_1967,scully_quantum_1997}.
Namely, we assume that the atom decays very fast and quickly arrives
at its steady state ($\kappa\ll\gamma_{h,c}$). That gives a set of
algebraic equations, which enable us to obtain the following equation
for the photon number probability $P_{n}:=\langle n|\varrho|n\rangle$
(see Appendix \ref{apx:elimination}), 
\begin{align}
\frac{d}{dt}P_{n}= & \frac{n\left[\mathscr{A}P_{n-1}-\mathscr{A}_{\mathrm{b}}P_{n}\right]}{1+n\mathscr{B}/\mathscr{A}}-\frac{(n+1)\left[\mathscr{A}P_{n}-\mathscr{A}_{\mathrm{b}}P_{n+1}\right]}{1+(n+1)\mathscr{B}/\mathscr{A}}\nonumber \\
 & +\kappa\left[(n+1)P_{n+1}-nP_{n}\right],\label{eq:laser}
\end{align}
where we denote 
\begin{gather}
\mathscr{A}:=\frac{4g^{2}\overline{\mathsf{n}}_{h}(\overline{\mathsf{n}}_{c}+1)}{\boldsymbol{\varGamma}\boldsymbol{\Phi}},\quad\mathscr{A}_{\mathrm{b}}=\frac{4g^{2}\overline{\mathsf{n}}_{c}(\overline{\mathsf{n}}_{h}+1)}{\boldsymbol{\varGamma}\boldsymbol{\Phi}},\nonumber \\
\mathscr{B}:=\mathscr{A}\cdot\frac{4g^{2}\boldsymbol{\Psi}}{\boldsymbol{\varGamma}\boldsymbol{\Phi}}.
\end{gather}
The constants $\boldsymbol{\varGamma}$, $\boldsymbol{\Phi}$, $\boldsymbol{\Psi}$
are the same as in Eqs.\,(\ref{eq:Gamma}, \ref{eq:dN}). This equation
has the same form as Ref.\,\cite{sargent_laser_1978} {[}eq.\,(59)
in pp.\,297{]}. Here $\mathscr{A}$ indicates the stimulated emission
rate, while $\mathscr{A}_{\mathrm{b}}$ is the stimulated absorption
rate. 

Expending the fractions in the above lasing equation to the 1st order,
we further derive the equation for the average photon number $\langle\hat{n}_{l}\rangle=\sum nP_{n}$,
i.e.,
\begin{align}
\frac{d}{dt}\langle\hat{n}_{l}\rangle= & \big(\mathscr{A}-\mathscr{A}_{\mathrm{b}}-\kappa\big)\langle\hat{n}_{l}\rangle\nonumber \\
 & +\mathscr{A}-\mathscr{B}\langle(\hat{n}_{l}+1)^{2}\rangle+\frac{\mathscr{A}}{\mathscr{A}_{\mathrm{b}}}\cdot\mathscr{B}\langle\hat{n}_{l}^{2}\rangle\dots
\end{align}
The first linear term is the net lasing gain, which is exactly the
same with that in the previous semi-classical laser equation (\ref{eq:da/dt}),
and we can verify $\mathscr{A}-\mathscr{A}_{\mathrm{b}}=\mathbf{G}$.
The $\mathscr{B}$ terms are nonlinear saturation which is beyond
the linearized laser theory.

\section{Photon number statistics}

Setting $\dot{P}_{n}=0$ in the lasing equation (\ref{eq:laser}),
the photon number distribution of the cavity mode in the steady state
is obtained as follows:
\begin{align}
\frac{P_{n}}{P_{n-1}} & =\frac{\mathscr{A}}{\mathscr{A}_{\mathrm{b}}+\kappa(1+\frac{n\mathscr{B}}{\mathscr{A}})},\label{eq:ratio}\\
P_{n} & =P_{0}\prod_{k=1}^{n}\frac{\mathscr{A}}{\mathscr{A}_{\mathrm{b}}+\kappa(1+\frac{k\mathscr{B}}{\mathscr{A}})}.\quad(n\ge1)\nonumber 
\end{align}
The maximum probability of $P_{n}$ appears around
\begin{equation}
n_{*}=\frac{\mathscr{A}}{\kappa\mathscr{B}}(\mathscr{A}-\mathscr{A}_{\mathrm{b}}-\kappa).\label{eq:maxima}
\end{equation}
 $P_{n}$ increases when $n<n_{*}$ while decreases when $n>n_{*}$.
Thus the lasing threshold requires $n_{*}\ge0$, which is just the
same as the above threshold condition $\mathbf{G}-\kappa=\mathscr{A}-\mathscr{A}_{\mathrm{b}}-\kappa\ge0$. 

When the system is working far below the threshold, approximately
the distribution becomes an exponentially decaying one,
\begin{gather}
\frac{P_{n}}{P_{n-1}}=\frac{\mathscr{A}}{\mathscr{A}_{\mathrm{b}}+\kappa}\le1.
\end{gather}
Therefore, the output light is like thermal radiation.

But we should remember if the system is below but still close to the
lasing threshold, the the realistic photon distribution is not the
idealistic thermal one {[}Eq.\,(\ref{eq:ratio}){]}. For example,
Fig.\,\ref{fig-wigner}(b, c) shows that $P_{n}$ is not exactly
an exponentially decaying distribution. As well, above the threshold,
the distribution is not the perfect Poisson one either \cite{scully_quantum_1967,scully_quantum_1997}.

In Fig.\,\ref{fig-wigner}, we show the photon number distributions
and the corresponding Wigner functions when $T_{h}$ is in different
regimes. The photon number distribution is calculated by the above
analytical result Eq.\,(\ref{eq:ratio}) (yellow columns on the right),
as well as by solving the master equation (\ref{eq:ME}) numerically
(green columns on the left), and they match each other quite well
for all different $T_{h}$, which confirms the validity of the above
adiabatic elimination method.

And it shows that with the increasing of  $T_{h}$, the cavity output
light first gives out thermal light, then becomes lasing, and turns
back to be thermal again at the very high temperature regime, which
confirms our previous result.

It is worth noticing that the two blue points in Fig.\,\ref{fig-threshold}(a)
($\overline{\mathsf{n}}_{h}\simeq0.507$ and $\overline{\mathsf{n}}_{h}\simeq2.629$)
have the same gain $\mathbf{G}$, but their distributions still differ
a lot {[}see Fig.\,\ref{fig-wigner}(c, e){]}. For example, their
maximum value also depends on $\mathscr{A}/\mathscr{B}$ {[}see Eq.\,(\ref{eq:maxima}){]}.

The total output power of the cavity is 
\begin{equation}
{\cal P}_{l}=-\mathrm{tr}\big[\mathcal{L}_{\mathsf{cav}}[\rho]\cdot\hbar\Omega_{l}\hat{n}_{l}\big]=\hbar\Omega_{l}\cdot\kappa\langle\hat{n}_{l}\rangle,
\end{equation}
which is proportional to the average photon number of the cavity mode.
From the photon number distribution Eq.\,(\ref{eq:ratio}), we obtain
the average photon number (see Appendix \ref{apx:elimination})
\begin{equation}
\langle\hat{n}_{l}\rangle=\frac{\mathscr{A}}{\kappa\mathscr{B}}(\mathscr{A}-\mathscr{A}_{\mathrm{b}}-\kappa)+\frac{\mathscr{A}}{\kappa\mathscr{B}}(\kappa+\mathscr{A}_{\mathrm{b}})P_{0}.\label{eq:N_L}
\end{equation}
 In Fig.\,\ref{fig-threshold}(c), we compare this analytical result
for cavity photon number with the numerical result by solving the
master equation (\ref{eq:ME}) directly, and they fit each other quite
well.

When the system is far above the threshold, $P_{0}\simeq0$, thus
only the first term dominates. Therefore, the laser power is 
\begin{align}
{\cal P}_{l} & =\hbar\Omega_{l}\cdot\frac{\mathscr{A}}{\mathscr{B}}(\mathscr{A}-\mathscr{A}_{\mathrm{b}}-\kappa)\nonumber \\
 & =\frac{\hbar\Omega_{l}\cdot\gamma_{h}\gamma_{c}(\overline{\mathsf{n}}_{h}-\overline{\mathsf{n}}_{c}-\frac{\kappa}{4g^{2}}\boldsymbol{\varGamma}\boldsymbol{\Phi})}{\gamma_{h}(3\overline{\mathsf{n}}_{h}+1)+\gamma_{c}(3\overline{\mathsf{n}}_{c}+1)}.\label{eq:power}
\end{align}
The leading term of this result (without the $\kappa$ term) is the
same with that in Ref.\,\cite{scully_quantum_2011}, which was calculated
by rate equations (see eq.\,{[}S6{]} in supporting information).
This result is valid when the system is far above the lasing threshold.
When the system is below or around the threshold, the $P_{0}$ term
in Eq.\,(\ref{eq:N_L}) becomes important and cannot be neglected
{[}Fig.\,\ref{fig-threshold}(c){]}. Considering $\overline{\mathsf{n}}_{c}\sim0$,
$\gamma_{h}=\gamma_{c}=\gamma$, a rough estimation for the cavity
photon number is 
\begin{equation}
\langle\hat{n}_{l}\rangle\sim\frac{\gamma\overline{\mathsf{n}}_{h}}{\kappa(3\overline{\mathsf{n}}_{h}+2)}-\frac{\gamma^{2}}{4g^{2}}\cdot\frac{(\overline{\mathsf{n}}_{h}+1)(2\overline{\mathsf{n}}_{h}+1)}{3\overline{\mathsf{n}}_{h}+2},\label{eq:estimationN}
\end{equation}
where the second term increases with $\overline{\mathsf{n}}_{h}$
monotonically, and indicates the hot photon number could weaken the
lasing. Thus the maximum cavity photon number does not appear at $\overline{\mathsf{n}}_{h}\rightarrow\infty$.
Again we see that the cavity photon number is not large, and this
is because there is only one single atom in the cavity, thus the photon
emission is limited.

Further, with this distribution $P_{n}$, the variance of the photon
number is
\begin{equation}
\sigma^{2}:=\langle\hat{n}_{l}^{2}\rangle-\langle\hat{n}_{l}\rangle^{2}=\frac{\mathscr{A}^{2}}{\kappa\mathscr{B}}-\frac{\mathscr{A}}{\kappa\mathscr{B}}(\kappa+\mathscr{A}_{\mathrm{b}})P_{0}\langle\hat{n}_{l}\rangle.
\end{equation}
When the system is far above the threshold, $P_{0}\simeq0$, and 
\begin{equation}
\frac{\sigma^{2}}{\langle\hat{n}_{l}\rangle}=1+\frac{\mathscr{A}_{\mathrm{b}}+\kappa}{\mathscr{A}-\mathscr{A}_{\mathrm{b}}-\kappa}.
\end{equation}
Thus the lasing photon number distribution is super-Poissonian ($\sigma^{2}>\langle\hat{n}_{l}\rangle$).
When $\mathscr{A}\gg\mathscr{A}_{\mathrm{b}}+\kappa$, the photon
distribution approaches the Poissonian one with $\sigma^{2}\simeq\langle\hat{n}_{l}\rangle$.

\section{Four-level heat engine model}

\begin{figure}
\includegraphics[width=0.5\columnwidth]{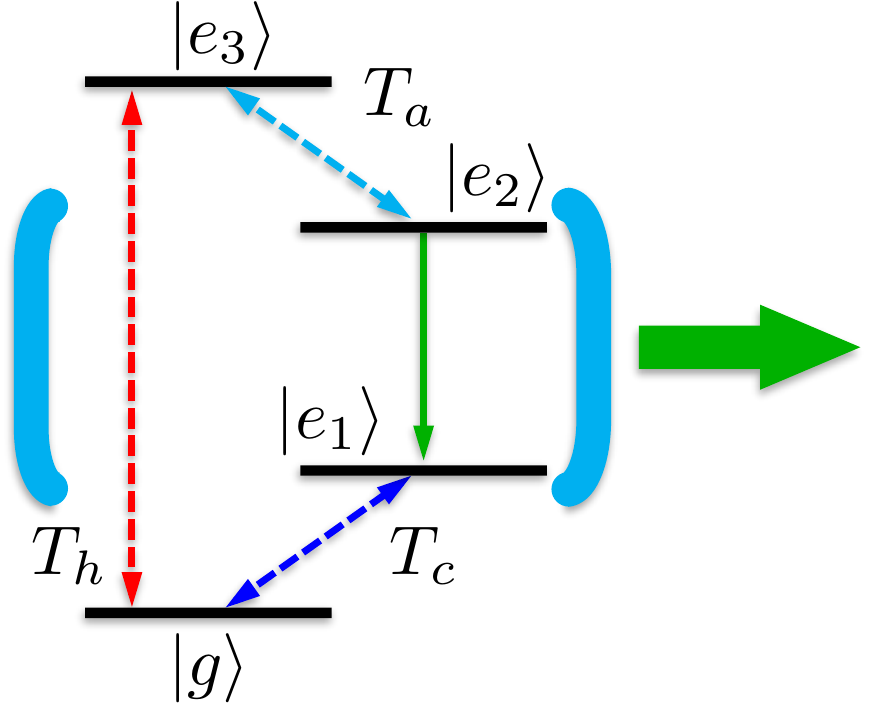}

\caption{(Color online) A four-level heat engine. The transition $|e_{2}\rangle\leftrightarrow|e_{3}\rangle$
is coupled with a third ancilla bath with temperature $T_{\mathsf{a}}$.}
\label{fig-4level}
\end{figure}

In the above discussion, we notice that the three-level heat engine
has a problem of double critical points, namely, when the hot bath
temperature is increased, the atomic coherence decay rate is also
increased, which decreases the lasing gain and even below the threshold
again. To avoid this problem, we consider a four-level system as shown
in Fig.\,\ref{fig-4level} \cite{yu_four-level_2010}. The transition
$|e_{2}\rangle\leftrightarrow|e_{3}\rangle$ is coupled with a third
ancilla bath with a low temperature $T_{\mathsf{a}}$, so as to ``cool
down'' the atomic coherence decay rate of the lasing transition.
Besides, this third bath also increases the population on $|e_{2}\rangle$,
as we will show below. 

Using the same method as the above discussion (see also Appendix \ref{apx:4-level}),
the linearized semi-classical lasing equation is 
\begin{gather}
\dot{{\cal E}}\simeq\frac{1}{2}\big[\cfrac{4g^{2}}{\boldsymbol{\varGamma}'}\cdot\Delta\mathrm{N}_{0}'-\kappa\big]{\cal E}+o(|{\cal E}|^{2}),\nonumber \\
\Delta\mathrm{N}_{0}'=[(\overline{\mathsf{n}}_{h}-\overline{\mathsf{n}}_{c})\overline{\mathsf{n}}_{\mathsf{a}}+(\overline{\mathsf{n}}_{c}+1)\overline{\mathsf{n}}_{h}]/\boldsymbol{\Phi}',\label{eq:lasing-4l}
\end{gather}
where $\mathbf{G}':=4g^{2}\Delta\mathrm{N}_{0}'/\boldsymbol{\varGamma}'$
is the lasing gain, and
\begin{align}
\boldsymbol{\varGamma}'= & \gamma_{\mathsf{a}}\overline{\mathsf{n}}_{\mathsf{a}}+\gamma_{c}(\overline{\mathsf{n}}_{c}+1),\nonumber \\
\boldsymbol{\Phi}'= & (4\overline{\mathsf{n}}_{h}\overline{\mathsf{n}}_{c}+3\overline{\mathsf{n}}_{h}+2\overline{\mathsf{n}}_{c}+1)\overline{\mathsf{n}}_{\mathsf{a}}+\overline{\mathsf{n}}_{h}(\overline{\mathsf{n}}_{c}+1).
\end{align}
Here $\overline{\mathsf{n}}_{\mathsf{a}}:=\overline{\mathrm{n}}_{\mathrm{\textsc{p}}}(\omega_{\mathsf{a}},T_{\mathsf{a}})$
is the thermal photon number of the transition $|e_{2}\rangle\leftrightarrow|e_{3}\rangle$,
and $\omega_{\mathsf{a}}=E_{3}-E_{2}$.

\begin{figure}
\includegraphics[width=1\columnwidth]{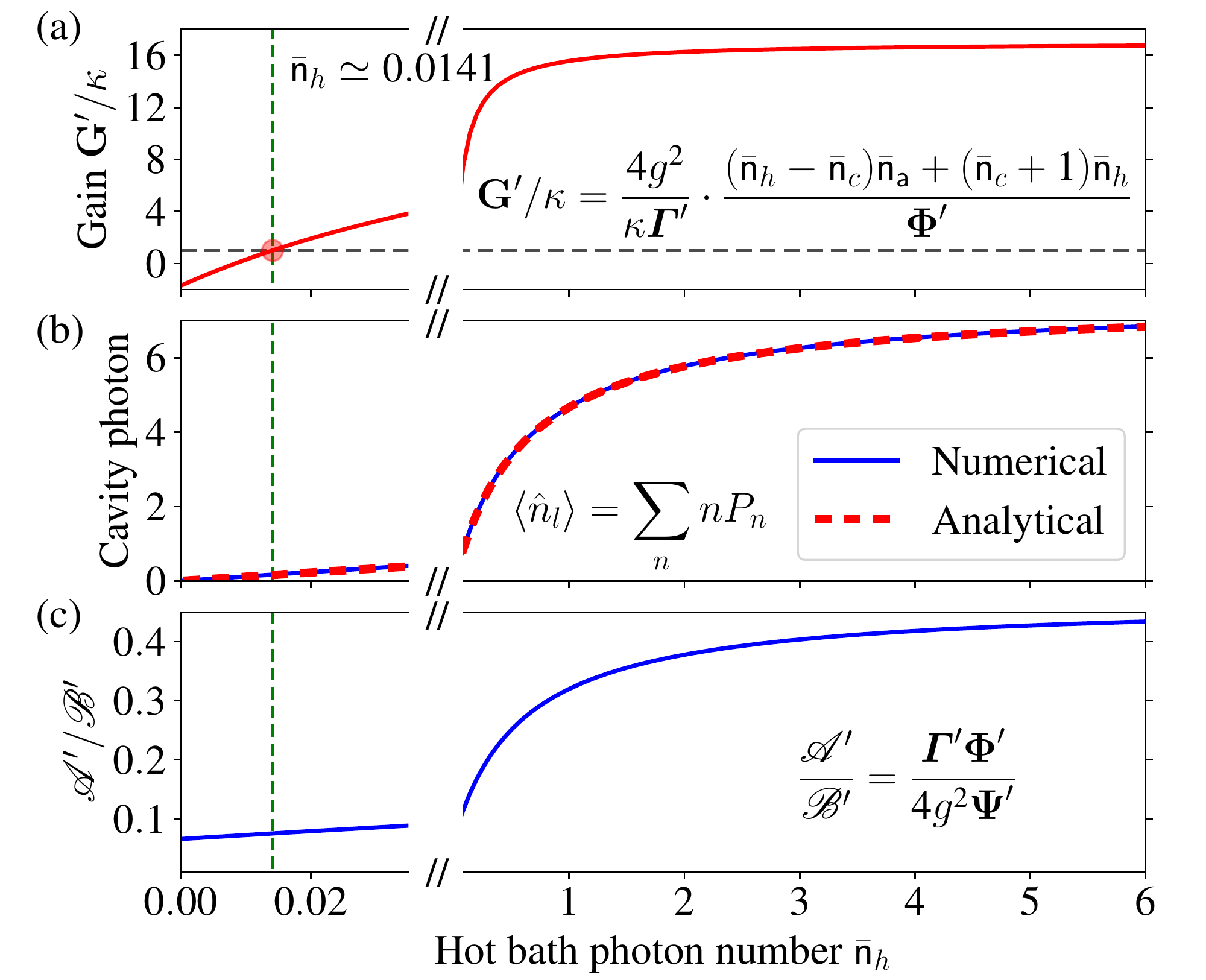}

\caption{(Color online) (a) The lasing gain $\mathbf{G}'$ for the four-level
system. (b) The average photon number $\langle\hat{n}_{l}\rangle$
in the cavity obtained from the analytical result (dashed red) and
numerically solving the master equation directly (solid blue). (c)
The ratio $\mathscr{A}'/\mathscr{B}'$. We set $\gamma_{h}=\gamma_{c}=\gamma_{\mathsf{a}}=32\kappa$,
$g=14\kappa$, and $\overline{\mathsf{n}}_{c}=0.1$, $\overline{\mathsf{n}}_{\mathsf{a}}=0.1$.
The critical point is $\overline{\mathsf{n}}_{h}\simeq0.0141$.}
\label{fig-4level-threshold}
\end{figure}

In this case, the decay rate $\boldsymbol{\varGamma}'$ does not depend
on the hot bath, thus will not increase with $T_{h}$ as the three-level
case. And it is clear to see that $\mathbf{G}'/\kappa=1$ is a linear
equation and gives only one root for $\overline{\mathsf{n}}_{h}$
when $\overline{\mathsf{n}}_{c,\mathsf{a}}$ are fixed, which means
only one critical point exists (see Fig.\,\ref{fig-4level-threshold}).

Simple algebra shows that $\Delta\mathrm{N}_{0}'$ is the population
inversion on $|e_{2}\rangle$ and $|e_{1}\rangle$ when there is no
cavity coupling. Notice that when $T_{\mathsf{a}}\rightarrow0$, we
have $\Delta\mathrm{N}_{0}'\rightarrow1$, which means all the populations
would fall on $|e_{2}\rangle$ in the steady state. This is because
when $T_{\mathsf{a}}=0$, once the population falls down from $|e_{3}\rangle$
to $|e_{2}\rangle$, it could never go back. This is the maximum inversion
for lasing. In Fig.\,\ref{fig-4level-threshold}, we also notice
that the lasing threshold is much easier to achieve comparing with
the 3-level case, i.e., a very small $\overline{\mathsf{n}}_{h}$
provides a strong enough pumping for lasing.

In the finite cavity limit, $\kappa\rightarrow0$, the lasing condition
is given by $\Delta\mathrm{N}_{0}'\ge0$ {[}Eq.\,(\ref{eq:lasing-4l}){]},
which leads to 
\begin{equation}
e^{\frac{\omega_{\mathsf{a}}}{T_{\mathsf{a}}}}\cdot e^{\frac{\omega_{c}}{T_{c}}}\ge e^{\frac{\omega_{h}}{T_{h}}}.
\end{equation}
If we consider the ancilla bath has the same temperature with the
cold one, $T_{\mathsf{a}}=T_{c}$, the above inequality gives 
\begin{equation}
1-\frac{T_{c}}{T_{h}}\ge1-\frac{\omega_{\mathsf{a}}+\omega_{c}}{\omega_{h}}=\frac{\Omega_{l}}{\omega_{h}}.
\end{equation}
Here $\Omega_{l}/\omega_{h}$ is just the output efficiency of this
four-level system, and again it is bounded by the Carnot efficiency,
which is similar as the previous SSDB discussion. 

The full-quantum equation also has the same form as the three-level
case {[}Eq.\,(\ref{eq:laser}){]}, except the parameters $\mathscr{A}$,
$\mathscr{A}_{\mathrm{b}}$, $\mathscr{B}$ should be changed to be
(see Appendix \ref{apx:4-level})
\begin{gather}
\mathscr{A}':=\frac{4g^{2}\overline{\mathsf{n}}_{h}(\overline{\mathsf{n}}_{c}+1)(\overline{\mathsf{n}}_{\mathsf{a}}+1)}{\boldsymbol{\varGamma}'\boldsymbol{\Phi}'},\quad\mathscr{A}_{\mathrm{b}}'=\frac{4g^{2}\overline{\mathsf{n}}_{c}\overline{\mathsf{n}}_{\mathsf{a}}(\overline{\mathsf{n}}_{h}+1)}{\boldsymbol{\varGamma}'\boldsymbol{\Phi}'},\nonumber \\
\mathscr{B}':=\mathscr{A}'\cdot\frac{4g^{2}\boldsymbol{\Psi}'}{\boldsymbol{\varGamma}'\boldsymbol{\Phi}'},
\end{gather}
where $\boldsymbol{\Psi}'=\gamma_{h}^{-1}(4\overline{\mathsf{n}}_{\mathsf{a}}\overline{\mathsf{n}}_{c}+\overline{\mathsf{n}}_{\mathsf{a}}+3\overline{\mathsf{n}}_{c}+1)+\gamma_{c}^{-1}(4\overline{\mathsf{n}}_{h}\overline{\mathsf{n}}_{\mathsf{a}}+2\overline{\mathsf{n}}_{h}+\overline{\mathsf{n}}_{\mathsf{a}})+\gamma_{\mathsf{a}}^{-1}(4\overline{\mathsf{n}}_{h}\overline{\mathsf{n}}_{c}+2\overline{\mathsf{n}}_{h}+3\overline{\mathsf{n}}_{c}+1)$.
In Fig.\,\ref{fig-4level-threshold} we show the lasing gain and
the cavity photon number, and they all increases monotonically with
the hot bath temperature $T_{h}$. Again, the laser gain is just given
by $\mathbf{G}'=\mathscr{A}'-\mathscr{A}_{\mathrm{b}}'$.

The cavity photon number is still given by Eq.\,(\ref{eq:N_L}),
but the parameters should be changed by $\mathscr{A}',\,\mathscr{A}_{\mathrm{b}}'$
and $\mathbf{\mathscr{B}'}$ correspondingly. Fig.\,\ref{fig-4level-threshold}(c)
shows that this analytical result for the cavity photon number fits
quite well with the numerical result.

When the system is far above threshold, the laser power is estimated
by (considering $\kappa\rightarrow0$)
\begin{equation}
\kappa\langle\hat{n}_{l}\rangle\simeq\frac{\mathbf{G}'\mathscr{A}'}{\mathscr{B}'}=[(\overline{\mathsf{n}}_{h}-\overline{\mathsf{n}}_{c})\overline{\mathsf{n}}_{\mathsf{a}}+(\overline{\mathsf{n}}_{c}+1)\overline{\mathsf{n}}_{h}]/\boldsymbol{\Psi}'.
\end{equation}
 If we further consider $\overline{\mathsf{n}}_{c,\mathsf{a}}\sim0$,
$\gamma_{h}=\gamma_{c}=\gamma_{\mathsf{a}}=\gamma$, $\overline{\mathsf{n}}_{h}\gg1$,
then the maximum gain and cavity photon number are around $\mathbf{G}'\sim4g^{2}/\gamma$
and $\langle\hat{n}_{l}\rangle\sim\gamma/4\kappa$. Both the lasing
gain $\mathbf{G}'$ and the cavity photon number $\langle\hat{n}_{l}\rangle$
approach saturated at the very high temperature regime, as shown in
Fig.\,\ref{fig-4level-threshold}. This is because in this regime,
the population has been almost totally inverted ($\Delta\mathrm{N}_{0}'\rightarrow1$),
thus the increase of the hot bath temperature $T_{h}$ can no longer
bring in a significant increase to the lasing gain. Unlike the 3-level
result Eq.\,(\ref{eq:estimationN}), the hot bath no longer has any
weakening effect to the lasing, thus more lasing photons can be produced
in the cavity, and the lasing power can be increased. But still the
cavity photon number is limited due to the single atom feature.

\section{Summary}

In this paper, we study the statistics of the lasing output from the
SSDB heat engine. In this heat engine model, a single three-level
atom is coupled with the quantized cavity mode, as well as contacting
with a hot and a cold heat bath together. We derive a laser equation
for this heat engine model, and obtain the photon number distribution
for both below and above the lasing threshold. Below the lasing threshold,
the output light from the cavity is more likely thermal radiation.
With the increase of the hot bath temperature, the population is inverted
and lasing light comes out. If the hot bath temperature keeps increasing,
our analytical result show that the atomic decay rate is also enhanced,
which weakens the lasing gain. As the result, at a very high temperature
of the hot bath, another critical point appears, and after that the
output light become thermal radiation again. 

To avoid this double-threshold behavior, we considered a four-level
model where neither of the two lasing level is coupled with the hot
bath directly, and a third ancilla bath is introduced. As the result,
the atomic decay rate in this four-level no longer depends on the
hot bath temperature, and thus the lasing gain and cavity photon number
keeps increasing monotonically when the hot bath temperature increases.
This four-level heat engine is also bounded by the Carnot efficiency,
which is the same as the original three-level SSDB model.

\emph{Acknowledgement} - This study is supported by Office of Naval
Research (Award No. N00014-16-1-3054) and Robert A. Welch Foundation
(Grant No. A-1261).

\appendix
\begin{widetext}

\section{Lasing equation for the three-level system \label{apx:elimination}}

\noindent\textbf{ 1. Lasing equation: }\vspace{.5em}

Here we derive the lasing equation for the photon number distribution
$P_{n}=\langle n|\varrho|n\rangle$ where $\varrho=\mathrm{tr}_{\mathrm{atom}}\rho$
is the density matrix of the cavity mode. We assume the cavity leaking
is much slower than the atom decay and omit ${\cal L}_{\mathsf{cav}}[\rho]$,
then the master equation (\ref{eq:ME}) gives (denoting $\rho_{\alpha\beta;mn}=\langle\alpha,m|\rho|\beta,n\rangle$
where $\alpha,\beta=1,2,g$ is the atom state indices)
\begin{align}
\frac{d}{dt}\rho_{11;mn} & =ig\left(\sqrt{n}\rho_{12;m,n-1}-\sqrt{m}\rho_{21;m-1,n}\right)-\Gamma_{c}^{-}\rho_{11;mn}+\Gamma_{c}^{+}\rho_{gg;mn},\nonumber \\
\frac{d}{dt}\rho_{22;mn} & =ig\left(\sqrt{n+1}\rho_{21;m,n+1}-\sqrt{m+1}\rho_{12;m+1,n}\right)-\Gamma_{h}^{-}\rho_{22;mn}+\Gamma_{h}^{+}\rho_{gg;mn},\nonumber \\
\frac{d}{dt}\rho_{12;mn} & =ig\left(\sqrt{n+1}\rho_{11;m,n+1}-\sqrt{m}\rho_{22;m-1,n}\right)-\frac{1}{2}(\Gamma_{h}^{-}+\Gamma_{c}^{-})\rho_{12;mn},\label{eq:Rho_ij;mn}\\
\frac{d}{dt}\rho_{21;mn} & =-ig\left(\sqrt{m+1}\rho_{11;m+1,n}-\sqrt{n}\rho_{22;m,n-1}\right)-\frac{1}{2}(\Gamma_{h}^{-}+\Gamma_{c}^{-})\rho_{21;mn},\nonumber \\
\frac{d}{dt}\rho_{gg;mn} & =\Gamma_{c}^{-}\rho_{11;mn}-\Gamma_{c}^{+}\rho_{gg;mn}+\Gamma_{h}^{-}\rho_{22;mn}-\Gamma_{h}^{+}\rho_{gg;mn}.\nonumber 
\end{align}
Here we denote $\Gamma_{i}^{+}=\gamma_{i}\overline{\mathsf{n}}_{i}$
and $\Gamma_{i}^{-}=\gamma_{i}(\overline{\mathsf{n}}_{i}+1)$ for
$i=h,c$. The matrix elements for the cavity mode is $P_{mn}:=\langle m|\varrho|n\rangle=\rho_{11;mn}+\rho_{22;mn}+\rho_{gg;mn}$,
thus, combining with the cavity leaking term ${\cal L}_{\mathsf{cav}}[\rho]$,
the equation for the cavity mode is
\begin{align}
\frac{d}{dt}P_{mn}= & ig\big(\sqrt{n}\rho_{12;m,n-1}-\sqrt{m}\rho_{21;m-1,n}\big)-ig\big(\sqrt{m+1}\rho_{12;m+1,n}-\sqrt{n+1}\rho_{21;m,n+1}\big)\nonumber \\
 & +\kappa\big[\sqrt{(m+1)(n+1)}P_{m+1,n+1}-\frac{1}{2}(m+n)P_{mn}\big].\label{eq:rho_a}
\end{align}
In the first two terms, the dynamics of the cavity mode is still coupled
with that of the atom. 

To derive a equation for the cavity mode alone, we need to replace
$\rho_{12;mn}$ by $P_{mn}$ in the above equation by adiabatic elimination
\cite{scully_quantum_1967,scully_quantum_1997}. That is, due to the
fast decay of the atom, Eq.\,(\ref{eq:Rho_ij;mn}) quickly arrives
at the steady state, and that gives:
\begin{align}
0 & =ig\left(\sqrt{n}\rho_{12;m,n-1}-\sqrt{m}\rho_{21;m-1,n}\right)-\Gamma_{c}^{-}\rho_{11;mn}+\Gamma_{c}^{+}\rho_{gg;mn},\nonumber \\
0 & =ig\left(\sqrt{n}\rho_{21;m-1,n}-\sqrt{m}\rho_{12;m,n-1}\right)-\Gamma_{h}^{-}\rho_{22;m-1,n-1}+\Gamma_{h}^{+}\rho_{gg;m-1,n-1},\nonumber \\
0 & =ig\left(\sqrt{n}\rho_{11;mn}-\sqrt{m}\rho_{22;m-1,n-1}\right)-\frac{1}{2}(\Gamma_{h}^{-}+\Gamma_{c}^{-})\rho_{12;m,n-1},\nonumber \\
0 & =-ig\left(\sqrt{m}\rho_{11;mn}-\sqrt{n}\rho_{22;m-1,n-1}\right)-\frac{1}{2}(\Gamma_{h}^{-}+\Gamma_{c}^{-})\rho_{21;m-1,n},\\
0 & =\Gamma_{c}^{-}\rho_{11;mn}-\Gamma_{c}^{+}\rho_{gg;mn}+\Gamma_{h}^{-}\rho_{22;mn}-\Gamma_{h}^{+}\rho_{gg;mn},\nonumber \\
0 & =\Gamma_{c}^{-}\rho_{11;m-1,n-1}-\Gamma_{c}^{+}\rho_{gg;m-1,n-1}+\Gamma_{h}^{-}\rho_{22;m-1,n-1}-\Gamma_{h}^{+}\rho_{gg;m-1,n-1}.\nonumber 
\end{align}
Together with the relations
\begin{align}
P_{mn} & =\rho_{11;mn}+\rho_{22;mn}+\rho_{gg;mn},\\
P_{m-1,n-1} & =\rho_{11;m-1,n-1}+\rho_{22;m-1,n-1}+\rho_{gg;m-1,n-1},\nonumber 
\end{align}
these equations becomes a closed set for the 8 variables $\rho_{gg;mn}$,
$\rho_{11;mn}$, $\rho_{22;mn}$, $\rho_{gg;m-1,n-1}$, $\rho_{11;m-1,n-1}$,
$\rho_{22;m-1,n-1}$, $\rho_{12;m,n-1}$, $\rho_{21;m-1,n}$. Solving
this equation set, we obtain the steady values of $\rho_{12;mn}$
represented by $P_{mn}$. Here we only concern about the diagonal
terms $P_{n}=\langle n|\varrho|n\rangle$ ($m=n$), and that gives
\begin{equation}
ig\left(\sqrt{n}\rho_{12;n,n-1}-\sqrt{n}\rho_{21;n-1,n}\right)=\frac{n\left[4g^{2}\overline{\mathsf{n}}_{h}(\overline{\mathsf{n}}_{c}+1)P_{n-1}-4g^{2}\overline{\mathsf{n}}_{c}(\overline{\mathsf{n}}_{h}+1)P_{n}\right]}{\boldsymbol{\varGamma}\boldsymbol{\Phi}+n\cdot4g^{2}\boldsymbol{\Psi}}
\end{equation}
for the first two terms in Eq.\,(\ref{eq:rho_a}), where 
\begin{equation}
\boldsymbol{\varGamma}:=\gamma_{c}(\overline{\mathsf{n}}_{c}+1)+\gamma_{h}(\overline{\mathsf{n}}_{h}+1),\qquad\boldsymbol{\Phi}=3\overline{\mathsf{n}}_{h}\overline{\mathsf{n}}_{c}+2(\overline{\mathsf{n}}_{h}+\overline{\mathsf{n}}_{c})+1,\qquad\boldsymbol{\Psi}:=\frac{1}{\gamma_{h}\gamma_{c}}[\gamma_{h}(3\overline{\mathsf{n}}_{h}+1)+\gamma_{c}(3\overline{\mathsf{n}}_{c}+1)],
\end{equation}
Then we obtain the lasing equation for the cavity mode {[}see eq.\,(59)
in pp.\,297 Ref.\,\cite{sargent_laser_1978}{]}
\begin{equation}
\frac{d}{dt}P_{n}=\frac{n\left[\mathscr{A}P_{n-1}-\mathscr{A}_{\mathrm{b}}P_{n}\right]}{1+n\mathscr{B}/\mathscr{A}}-\frac{(n+1)\left[\mathscr{A}P_{n}-\mathscr{A}_{\mathrm{b}}P_{n+1}\right]}{1+(n+1)\mathscr{B}/\mathscr{A}}+\kappa\left[(n+1)P_{n+1}-nP_{n}\right],\label{eq:laser-x}
\end{equation}
where we define
\begin{equation}
\mathscr{A}:=\frac{4g^{2}\overline{\mathsf{n}}_{h}(\overline{\mathsf{n}}_{c}+1)}{\boldsymbol{\varGamma}\boldsymbol{\Phi}},\qquad\mathscr{A}_{\mathrm{b}}:=\frac{4g^{2}\overline{\mathsf{n}}_{c}(\overline{\mathsf{n}}_{h}+1)}{\boldsymbol{\varGamma}\boldsymbol{\Phi}},\qquad\mathscr{B}:=\mathscr{A}\cdot\frac{4g^{2}\boldsymbol{\Psi}}{\boldsymbol{\varGamma}\boldsymbol{\Phi}}.
\end{equation}

\noindent\textbf{ 2. Photon number statistics: }\vspace{.5em}

In the above equation of $\dot{P}_{n}$, expending the fractions to
the 1st order, the average photon number $\langle\hat{n}_{l}\rangle=\sum nP_{n}$
gives 
\begin{equation}
\frac{d}{dt}\langle\hat{n}_{l}\rangle=\big(\mathscr{A}-\mathscr{A}_{\mathrm{b}}-\kappa\big)\langle\hat{n}_{l}\rangle+\mathscr{A}-\mathscr{B}\langle(\hat{n}_{l}+1)^{2}\rangle+\frac{\mathscr{A}}{\mathscr{A}_{\mathrm{b}}}\cdot\mathscr{B}\langle\hat{n}_{l}^{2}\rangle+\dots
\end{equation}
In the steady state, the photon number distribution is
\begin{equation}
\frac{P_{n}}{P_{n-1}}=\frac{\mathscr{A}}{\mathscr{A}_{\mathrm{b}}+\kappa(1+\frac{n\mathscr{B}}{\mathscr{A}})},\qquad P_{n}=P_{0}\prod_{k=1}^{n}\frac{(\mathscr{A}^{2}/\kappa\mathscr{B})}{\frac{\mathscr{A}}{\kappa\mathscr{B}}(\kappa+\mathscr{A}_{\mathrm{b}})+k}:=\frac{P_{0}Y!\,X^{n}}{(n+Y)!},
\end{equation}
where we define $X:=\mathscr{A}^{2}/\kappa\mathscr{B}$, $Y:=\frac{\mathscr{A}}{\kappa\mathscr{B}}(\kappa+\mathscr{A}_{\mathrm{b}})$.
The average photon number is
\begin{align}
\langle\hat{n}_{l}\rangle & =\sum_{n=0}^{\infty}n\cdot\frac{P_{0}Y!\,X^{n}}{(n+Y)!}=P_{0}Y!\cdot\sum_{n=1}^{\infty}\frac{(n+Y-Y)X^{n}}{(n+Y)!}=P_{0}Y!\cdot\sum_{n=1}^{\infty}\big[\frac{X\cdot X^{n-1}}{(n-1+Y)!}-\frac{YX^{n}}{(n+Y)!}\big]\nonumber \\
 & =X-Y+YP_{0}=\frac{\mathscr{A}}{\kappa\mathscr{B}}(\mathscr{A}-\mathscr{A}_{\mathrm{b}}-\kappa)+\frac{\mathscr{A}}{\kappa\mathscr{B}}(\kappa+\mathscr{A}_{\mathrm{b}})P_{0}.
\end{align}
When the system is far above the threshold, $P_{0}\simeq0$, then
we obtain
\begin{equation}
\kappa\langle\hat{n}_{l}\rangle=\frac{\mathscr{A}}{\mathscr{B}}(\mathscr{A}-\mathscr{A}_{\mathrm{b}}-\kappa)=\frac{\gamma_{h}\gamma_{c}(\overline{\mathsf{n}}_{h}-\overline{\mathsf{n}}_{c}-\frac{\kappa}{4g^{2}}\boldsymbol{\varGamma}\boldsymbol{\Phi})}{\gamma_{h}(3\overline{\mathsf{n}}_{h}+1)+\gamma_{c}(3\overline{\mathsf{n}}_{c}+1)}.
\end{equation}
Notice that the radiation power of the cavity is just  ${\cal P}_{l}=-\hbar\Omega_{l}\cdot\frac{d}{dt}\langle\hat{n}_{l}\rangle\big|_{\mathsf{cav}}=\hbar\Omega_{l}\cdot\kappa\langle\hat{n}_{l}\rangle$.
The leading term of this result is consistent with that in Ref.\,\cite{scully_quantum_2011}.

The variance of the photon number distribution is calculated by 
\begin{align}
\langle\hat{n}_{l}^{2}\rangle & =\sum_{n=0}^{\infty}n^{2}\cdot\frac{P_{0}Y!\,X^{n}}{(n+Y)!}=P_{0}Y!\cdot\sum_{n=1}^{\infty}\big[\frac{nX\cdot X^{n-1}}{(n-1+Y)!}-\frac{nYX^{n}}{(n+Y)!}\big]\nonumber \\
 & =\sum_{n=0}^{\infty}(n+1)X\cdot\frac{P_{0}Y!X^{n}}{(n+Y)!}-nY\cdot\frac{P_{0}Y!X^{n}}{(n+Y)!}=\langle\hat{n}_{l}+1\rangle X-\langle\hat{n}_{l}\rangle Y,\\
\sigma^{2}: & =\langle\hat{n}_{l}^{2}\rangle-\langle\hat{n}_{l}\rangle^{2}=X-YP_{0}(X-Y+YP_{0})=\frac{\mathscr{A}^{2}}{\kappa\mathscr{B}}-\frac{\mathscr{A}}{\kappa\mathscr{B}}(\kappa+\mathscr{A}_{\mathrm{b}})P_{0}\langle\hat{n}_{l}\rangle.\nonumber 
\end{align}
When the system is far above the threshold, $P_{0}\simeq0$, and we
have
\begin{equation}
\sigma^{2}=\frac{\mathscr{A}^{2}}{\kappa\mathscr{B}}=\langle\hat{n}_{l}\rangle+\frac{\mathscr{A}}{\kappa\mathscr{B}}(\mathscr{A}_{\mathrm{b}}+\kappa),\qquad\frac{\sigma^{2}}{\langle\hat{n}_{l}\rangle}=1+\frac{\mathscr{A}_{\mathrm{b}}+\kappa}{\mathscr{A}-\mathscr{A}_{\mathrm{b}}-\kappa}.
\end{equation}
If we have $\mathscr{A}\gg\mathscr{A}_{\mathrm{b}}+\kappa$, the photon
distribution well approaches the Poisson one with $\sigma^{2}\simeq\langle\hat{n}_{l}\rangle$.

\section{Lasing equation for the four-level system \label{apx:4-level}}

\noindent\textbf{ 1. Semi-classical lasing equation: }\vspace{.5em}

Here we study the lasing equation for the four-level model shown in
Fig.\,\ref{fig-4level}. First we consider the semi-classical equations
similar like Eq.\,(\ref{eq:semi-classical}), and we have 
\begin{align}
\frac{d}{dt}\langle\hat{\textsc{n}}_{1}\rangle= & \gamma_{c}\big[\overline{\mathsf{n}}_{c}\langle\hat{\textsc{n}}_{g}\rangle-(\overline{\mathsf{n}}_{c}+1)\langle\hat{\textsc{n}}_{1}\rangle\big]-ig\big[\langle\hat{\sigma}^{-}\rangle\langle\hat{a}^{\dagger}\rangle-\langle\hat{\sigma}^{+}\rangle\langle\hat{a}\rangle\big],\nonumber \\
\frac{d}{dt}\langle\hat{\textsc{n}}_{2}\rangle= & -\gamma_{\mathsf{a}}\big[\overline{\mathsf{n}}_{\mathsf{a}}\langle\hat{\textsc{n}}_{2}\rangle-(\overline{\mathsf{n}}_{\mathsf{a}}+1)\langle\hat{\textsc{n}}_{3}\rangle\big]+ig\big[\langle\hat{\sigma}^{-}\rangle\langle\hat{a}^{\dagger}\rangle-\langle\hat{\sigma}^{+}\rangle\langle\hat{a}\rangle\big],\nonumber \\
\frac{d}{dt}\langle\hat{\textsc{n}}_{3}\rangle= & \gamma_{h}\big[\overline{\mathsf{n}}_{h}\langle\hat{\textsc{n}}_{g}\rangle-(\overline{\mathsf{n}}_{h}+1)\langle\hat{\textsc{n}}_{3}\rangle\big]+\gamma_{\mathsf{a}}\big[\overline{\mathsf{n}}_{\mathsf{a}}\langle\hat{\textsc{n}}_{2}\rangle-(\overline{\mathsf{n}}_{\mathsf{a}}+1)\langle\hat{\textsc{n}}_{3}\rangle\big],\label{eq:4level-semi}\\
\frac{d}{dt}\langle\hat{\sigma}^{-}\rangle= & ig\langle\hat{\textsc{n}}_{2}-\hat{\textsc{n}}_{1}\rangle\langle\hat{a}\rangle-\frac{1}{2}\boldsymbol{\varGamma}'\langle\hat{\sigma}^{-}\rangle,\nonumber \\
\frac{d}{dt}\langle\hat{a}\rangle= & -\frac{\kappa}{2}\langle\hat{a}\rangle-ig\langle\hat{\sigma}^{-}\rangle,\nonumber 
\end{align}
where we denote $\boldsymbol{\varGamma}'=\gamma_{\mathsf{a}}\overline{\mathsf{n}}_{\mathsf{a}}+\gamma_{c}(\overline{\mathsf{n}}_{c}+1)$
for the coherence decay rate. The steady state gives the population
inversion as
\begin{align}
\langle\hat{\textsc{n}}_{2}-\hat{\textsc{n}}_{1}\rangle= & \cfrac{(\overline{\mathsf{n}}_{h}-\overline{\mathsf{n}}_{c})\overline{\mathsf{n}}_{\mathsf{a}}+(\overline{\mathsf{n}}_{c}+1)\overline{\mathsf{n}}_{h}}{\boldsymbol{\Phi}'+\frac{4g^{2}|{\cal E}|^{2}}{\boldsymbol{\varGamma}'}\boldsymbol{\Psi}'},\label{eq:4level-inversion}\\
\boldsymbol{\Phi}'= & 4\overline{\mathsf{n}}_{\mathsf{a}}\overline{\mathsf{n}}_{h}\overline{\mathsf{n}}_{c}+3\overline{\mathsf{n}}_{h}\overline{\mathsf{n}}_{\mathsf{a}}+2\overline{\mathsf{n}}_{\mathsf{a}}\overline{\mathsf{n}}_{c}+\overline{\mathsf{n}}_{h}\overline{\mathsf{n}}_{c}+\overline{\mathsf{n}}_{h}+\overline{\mathsf{n}}_{\mathsf{a}},\nonumber \\
\boldsymbol{\Psi}'= & \gamma_{h}^{-1}(4\overline{\mathsf{n}}_{\mathsf{a}}\overline{\mathsf{n}}_{c}+\overline{\mathsf{n}}_{\mathsf{a}}+3\overline{\mathsf{n}}_{c}+1)+\gamma_{c}^{-1}(4\overline{\mathsf{n}}_{h}\overline{\mathsf{n}}_{\mathsf{a}}+2\overline{\mathsf{n}}_{h}+\overline{\mathsf{n}}_{\mathsf{a}})+\gamma_{\mathsf{a}}^{-1}(4\overline{\mathsf{n}}_{h}\overline{\mathsf{n}}_{c}+2\overline{\mathsf{n}}_{h}+3\overline{\mathsf{n}}_{c}+1).\nonumber 
\end{align}
Therefore, the lasing equation is
\begin{align}
\dot{{\cal E}} & =\frac{2g^{2}}{\boldsymbol{\varGamma}'}\langle\hat{\textsc{n}}_{2}-\hat{\textsc{n}}_{1}\rangle{\cal E}-\frac{\kappa}{2}{\cal E}=\Big[\cfrac{2g^{2}[(\overline{\mathsf{n}}_{h}-\overline{\mathsf{n}}_{c})\overline{\mathsf{n}}_{\mathsf{a}}+(\overline{\mathsf{n}}_{c}+1)\overline{\mathsf{n}}_{h}]}{\boldsymbol{\varGamma}'\boldsymbol{\Phi}'+4g^{2}|{\cal E}|^{2}\boldsymbol{\Psi}'}-\frac{\kappa}{2}\Big]{\cal E}\nonumber \\
 & \simeq\frac{1}{2}\big[\cfrac{4g^{2}[(\overline{\mathsf{n}}_{h}-\overline{\mathsf{n}}_{c})\overline{\mathsf{n}}_{\mathsf{a}}+(\overline{\mathsf{n}}_{c}+1)\overline{\mathsf{n}}_{h}]}{\boldsymbol{\varGamma}'\boldsymbol{\Phi}'}-\kappa\big]{\cal E}.
\end{align}

\noindent\textbf{ 2. Full-quantum approach: }\vspace{.5em}

Now we consider the full-quantum approach. Similarly like Eq.\,(\ref{eq:Rho_ij;mn}),
the equations for the density elements are 
\begin{align}
\frac{d}{dt}\rho_{11;mn} & =ig\left(\sqrt{n}\rho_{12;m,n-1}-\sqrt{m}\rho_{21;m-1,n}\right)-\Gamma_{c}^{-}\rho_{11;mn}+\Gamma_{c}^{+}\rho_{gg;mn},\nonumber \\
\frac{d}{dt}\rho_{22;mn} & =ig\left(\sqrt{n+1}\rho_{21;m,n+1}-\sqrt{m+1}\rho_{12;m+1,n}\right)-\Gamma_{\mathsf{a}}^{+}\rho_{22;mn}+\Gamma_{\mathsf{a}}^{-}\rho_{33;mn},\nonumber \\
\frac{d}{dt}\rho_{12;mn} & =ig\left(\sqrt{n+1}\rho_{11;m,n+1}-\sqrt{m}\rho_{22;m-1,n}\right)-\frac{1}{2}(\Gamma_{\mathsf{a}}^{+}+\Gamma_{c}^{-})\rho_{12;mn},\nonumber \\
\frac{d}{dt}\rho_{21;mn} & =-ig\left(\sqrt{m+1}\rho_{11;m+1,n}-\sqrt{n}\rho_{22;m,n-1}\right)-\frac{1}{2}(\Gamma_{\mathsf{a}}^{+}+\Gamma_{c}^{-})\rho_{21;mn},\\
\frac{d}{dt}\rho_{gg;mn} & =\Gamma_{c}^{-}\rho_{11;mn}-\Gamma_{c}^{+}\rho_{gg;mn}+\Gamma_{h}^{-}\rho_{33;mn}-\Gamma_{h}^{+}\rho_{gg;mn}.\nonumber \\
\frac{d}{dt}\rho_{33;mn} & =\Gamma_{\mathsf{a}}^{+}\rho_{22;mn}-\Gamma_{\mathsf{a}}^{-}\rho_{33;mn}-\Gamma_{h}^{-}\rho_{33;mn}+\Gamma_{h}^{+}\rho_{gg;mn}.\nonumber 
\end{align}
Here we denote $\Gamma_{i}^{+}=\gamma_{i}\overline{\mathsf{n}}_{i}$
and $\Gamma_{i}^{-}=\gamma_{i}(\overline{\mathsf{n}}_{i}+1)$ for
$i=h,c,\mathsf{a}$. And the equation for the cavity mode is
\begin{align}
\frac{d}{dt}P_{mn}= & ig\big(\sqrt{n}\rho_{12;m,n-1}-\sqrt{m}\rho_{21;m-1,n}\big)-ig\big(\sqrt{m+1}\rho_{12;m+1,n}-\sqrt{n+1}\rho_{21;m,n+1}\big)\nonumber \\
 & +\kappa\big[\sqrt{(m+1)(n+1)}P_{m+1,n+1}-\frac{1}{2}(m+n)P_{mn}\big].
\end{align}
The first two terms mean cavity mode is coupled with the atom. 

We apply the adiabatic elimination, and consider the steady state
of the atom
\begin{align}
0 & =ig\left(\sqrt{n}\rho_{12;m,n-1}-\sqrt{m}\rho_{21;m-1,n}\right)-\Gamma_{c}^{-}\rho_{11;mn}+\Gamma_{c}^{+}\rho_{gg;mn},\nonumber \\
0 & =ig\left(\sqrt{n}\rho_{21;m-1,n}-\sqrt{m}\rho_{12;m,n-1}\right)-\Gamma_{\mathsf{a}}^{+}\rho_{22;m-1,n-1}+\Gamma_{\mathsf{a}}^{-}\rho_{33;m-1,n-1},\nonumber \\
0 & =ig\left(\sqrt{n}\rho_{11;mn}-\sqrt{m}\rho_{22;m-1,n-1}\right)-\frac{1}{2}(\Gamma_{\mathsf{a}}^{+}+\Gamma_{c}^{-})\rho_{12;m,n-1},\nonumber \\
0 & =-ig\left(\sqrt{m}\rho_{11;mn}-\sqrt{n}\rho_{22;m-1,n-1}\right)-\frac{1}{2}(\Gamma_{\mathsf{a}}^{+}+\Gamma_{c}^{-})\rho_{21;m-1,n},\nonumber \\
0 & =\Gamma_{c}^{-}\rho_{11;mn}-\Gamma_{c}^{+}\rho_{gg;mn}+\Gamma_{h}^{-}\rho_{33;mn}-\Gamma_{h}^{+}\rho_{gg;mn},\\
0 & =\Gamma_{\mathsf{a}}^{+}\rho_{22;mn}-\Gamma_{\mathsf{a}}^{-}\rho_{33;mn}-\Gamma_{h}^{-}\rho_{33;mn}+\Gamma_{h}^{+}\rho_{gg;mn},\nonumber \\
0 & =\Gamma_{c}^{-}\rho_{11;m-1,n-1}-\Gamma_{c}^{+}\rho_{gg;m-1,n-1}+\Gamma_{h}^{-}\rho_{33;m-1,n-1}-\Gamma_{h}^{+}\rho_{gg;m-1,n-1},\nonumber \\
0 & =\Gamma_{\mathsf{a}}^{+}\rho_{22;m-1,n-1}-\Gamma_{\mathsf{a}}^{-}\rho_{33;m-1,n-1}-\Gamma_{h}^{-}\rho_{33;m-1,n-1}+\Gamma_{h}^{+}\rho_{gg;m-1,n-1}.\nonumber 
\end{align}
Together with the relations
\begin{align}
P_{mn} & =\rho_{11;mn}+\rho_{22;mn}+\rho_{33;mn}+\rho_{gg;mn},\\
P_{m-1,n-1} & =\rho_{11;m-1,n-1}+\rho_{22;m-1,n-1}+\rho_{33;m-1,n-1}+\rho_{gg;m-1,n-1},\nonumber 
\end{align}
these equations becomes a closed set for the 10 variables $\rho_{gg;mn}$,
$\rho_{11;mn}$, $\rho_{22;mn}$, $\rho_{33;mn}$, $\rho_{gg;m-1,n-1}$,
$\rho_{11;m-1,n-1}$, $\rho_{22;m-1,n-1}$, $\rho_{33;m-1,n-1}$,
$\rho_{12;m,n-1}$, $\rho_{21;m-1,n}$. Solving this equation set,
we obtain 
\begin{equation}
ig\left(\sqrt{n}\rho_{12;n,n-1}-\sqrt{n}\rho_{21;n-1,n}\right)=\frac{4g^{2}n\left[\overline{\mathsf{n}}_{h}(\overline{\mathsf{n}}_{c}+1)(\overline{\mathsf{n}}_{\mathsf{a}}+1)P_{n-1}-\overline{\mathsf{n}}_{c}\overline{\mathsf{n}}_{\mathsf{a}}(\overline{\mathsf{n}}_{h}+1)P_{n}\right]}{\boldsymbol{\varGamma}'\boldsymbol{\Phi}'+n\cdot4g^{2}\boldsymbol{\Psi}'},
\end{equation}
where the parameters $\boldsymbol{\varGamma}'$, $\boldsymbol{\Phi}'$,
$\boldsymbol{\Psi}'$ are just the same as those in the semi-classical
results {[}Eqs.\,(\ref{eq:4level-semi}, \ref{eq:4level-inversion}){]}.
Thus, the laser equation has the same form as the three-level case
{[}Eqs.\,(\ref{eq:laser}, \ref{eq:laser-x}){]}, except the parameters
$\mathscr{A}$, $\mathscr{A}_{\mathrm{b}}$, $\mathscr{B}$ are changed
to be 
\begin{equation}
\mathscr{A}':=\frac{4g^{2}\overline{\mathsf{n}}_{h}(\overline{\mathsf{n}}_{c}+1)(\overline{\mathsf{n}}_{\mathsf{a}}+1)}{\boldsymbol{\varGamma}'\boldsymbol{\Phi}'},\qquad\mathscr{A}_{\mathrm{b}}'=\frac{4g^{2}\overline{\mathsf{n}}_{c}\overline{\mathsf{n}}_{\mathsf{a}}(\overline{\mathsf{n}}_{h}+1)}{\boldsymbol{\varGamma}'\boldsymbol{\Phi}'},\qquad\mathscr{B}':=\mathscr{A}'\cdot\frac{4g^{2}\boldsymbol{\Psi}'}{\boldsymbol{\varGamma}'\boldsymbol{\Phi}'}.
\end{equation}

\end{widetext}

\bibliographystyle{apsrev4-1}
\bibliography{Refs}

\end{document}